\documentclass[aps,preprint]{revtex4}%
\usepackage{amsfonts}
\usepackage{amsmath}
\usepackage{amssymb}
\usepackage{graphicx}%
\providecommand{\U}[1]{\protect\rule{.1in}{.1in}}

\begin{document}
\title[ ]{Classical Zero-Point Radiation and Relativity: The Problem of Blackbody
Radiation Revisited}
\author{Timothy H. Boyer}
\affiliation{Department of Physics, City College of the City University of New York, New
York, New York 10031}

\begin{abstract}
The physicists of the early 20th century were unaware of two ideas which are
vital to understanding some aspects of modern physics within classical theory.
\ The two ideas are: 1) the presence of classical electromagnetic zero-point
radiation, and 2) the importance of special relativity. \ In classes of modern
physics today, the problem of blackbody radiation within classical physics is
still described in the historical context of the early 20th century.
\ However, the inclusion of classical zero-point radiation and of relativity
now allows a completely satisfactory classical understanding of blackbody
radiation with the Planck spectrum, as well as of some other aspects of modern
physics. \ Here we sketch the current classical understanding of blackbody
radiation, pointing out that thermodynamics allows the presence of classical
zero-point radiation, and that use of \textit{nonrelativistic} physics leads
to the Rayleigh-Jeans spectrum while \textit{relativistic} physics gives the
Planck spectrum. \ The current textbooks of modern physics are a century out
of date in presenting the connections between classical and quantum physics.

\end{abstract}
\maketitle

\section{Introduction}

Current textbooks of modern physics are a century out-of-date in their
discussions of the relations between classical and quantum physics. \ The
textbooks still treat the connections in the historical context of the early
20th century.\cite{texts} \ Today it is well (but not widely) known that the
classical physicists of the early 20th century were unaware of two crucial
ideas vital to classical physics: 1) the presence of classical electromagnetic
zero-point radiation and 2) the importance of relativity. \ When these two
aspects are included, classical physics accounts for phenomena which the
current texts regard as exclusively quantum phenomena. \ These phenomena
include blackbody radiation, Casimir forces, van der Waals forces, harmonic
oscillator behavior, decrease of specific heats at low temperatures, and
diamagnetism.\cite{Rev1}\cite{Rev2}\cite{Rev3} \ In the present article, we
review how the two crucial missing aspects transform the classical
understanding of the blackbody radiation spectrum.

\section{Review of Thermal Radiation in Classical Physics}

\subsubsection{Radiation normal Modes}

Within classical physics, thermal radiation is treated as random classical
electromagnetic cavity radiation which is invariant under scattering.
\ Choosing for simplicity a rectangular, conducting-walled cavity of
dimensions $a,~b,~d$, the radiation inside can be written as a sum over the
radiation normal modes with vanishing scalar potential $\Phi$ and with vector
potential $\mathbf{A}$\ given by\cite{cavity}%
\begin{align}
\mathbf{A}(x,y,z,t) &  =%
{\textstyle\sum_{l,m,n=0}^{\infty}}
{\textstyle\sum_{\lambda=1}^{2}}
q_{lmn,\lambda}c\left(  \frac{32\pi}{abd}\right)  ^{1/2}\left\{
\widehat{i}\varepsilon_{lmnx}^{(\lambda)}\cos\left(  \frac{l\pi x}{a}\right)
\sin\left(  \frac{m\pi y}{b}\right)  \sin\left(  \frac{n\pi z}{d}\right)
\right.  \nonumber\\
&  +\widehat{j}\varepsilon_{lmny}^{(\lambda)}\sin\left(  \frac{l\pi x}%
{a}\right)  \cos\left(  \frac{m\pi y}{b}\right)  \sin\left(  \frac{n\pi z}%
{d}\right)  \nonumber\\
&  \left.  +\widehat{k}\varepsilon_{lmnz}^{(\lambda)}\sin\left(  \frac{l\pi
x}{a}\right)  \sin\left(  \frac{m\pi y}{b}\right)  \cos\left(  \frac{n\pi
z}{d}\right)  \right\}  \label{A1}%
\end{align}
where $\widehat{\varepsilon}_{lmn}^{(\lambda)}$ with $\lambda=1,2$ are the
mutually orthogonal unit vectors satisfying $\varepsilon_{x}l+\varepsilon
_{y}m+\varepsilon_{z}n=0$ , where $q_{lmn,\lambda}$ is the time-varying
amplitude of the mode, and where the frequency of the mode is given by
$\omega_{lmn}=c\pi(l^{2}/a^{2}+m^{2}/b^{2}+n^{2}/d^{2})^{1/2}.$ \ The
radiation energy in the box is given by $\mathcal{E}=[1/(8\pi)]%
{\textstyle\int}
{\textstyle\int}
{\textstyle\int}
dxdydz(E^{2}+B^{2})$ where $\mathbf{E}=-\nabla\Phi-(1/c)\partial
\mathbf{A}/\partial t$ and $\mathbf{B}=\nabla\times\mathbf{A},$ so that
\begin{equation}
\mathcal{E}=%
{\textstyle\sum_{l,m,n=0}^{\infty}}
{\textstyle\sum_{\lambda=1}^{2}}
(1/2)(\dot{q}_{lmn,\lambda}^{2}+\omega_{lmn,\lambda}^{2}q_{lmn,\lambda}^{2})=%
{\textstyle\sum_{l,m,n=0}^{\infty}}
{\textstyle\sum_{\lambda=1}^{2}}
J_{lmn,\lambda}\omega_{lmn,\lambda}\label{E}%
\end{equation}
where $J_{lmn,\lambda}$ is the action variable $J=%
{\textstyle\int}
p\,dq$ of the mode,\cite{Gold} $J_{lmn,\lambda}=\mathcal{E}_{lmn,\lambda
}/\omega_{lmn,\lambda},$ and $\mathcal{E}_{lmn,\lambda}$ is the energy of the
mode. \ Thus the energy of thermal radiation in a cavity can be expressed as a
sum over the energies of the normal modes of oscillation, with each mode
taking the form of a harmonic oscillator\cite{cavity2}%
\begin{equation}
\mathcal{E}=(1/2)(\dot{q}^{2}+\omega^{2}q^{2})\label{osc}%
\end{equation}

\subsubsection{Thermodynamics of the Simple Harmonic Oscillator}

Now the thermodynamics of a harmonic oscillator takes a particularly simple
form because the system has only two thermodynamic variables $T$ and $\omega
$.\cite{SHO} \ In thermal equilibrium with a bath, the average oscillator
energy is denoted by $U=\left\langle \mathcal{E}\right\rangle =\left\langle
J\right\rangle \omega,$ and satisfies $dQ=dU-dW$ with the entropy $S$
satisfying $dS=dQ/T.$ \ Now since $J$ is an adiabatic invariant,\cite{Gold}
the work done on the system is given by $dW=\left\langle J\right\rangle
d\omega=(U/\omega)d\omega.$ \ Combing these equations, we have
$dS=dQ/T=[dU-(U/\omega)d\omega]/T.$ $\ $Writing the differentials in terms of
$T$ and $\omega,$ we have $dS=(\partial S/T)dT+(\partial S/\partial
\omega)d\omega$ and $dU=(\partial U/dT)dT+(\partial U/\partial\omega)d\omega.$
\ Therefore $\partial S/\partial T=(\partial U/\partial T)/T$ and $\partial
S/\partial\omega=[(U/\omega)+(\partial U/\partial\omega)]/T.$ \ Now equating
the mixed second partial derivatives $\partial^{2}S/\partial T\partial
\omega=\partial^{2}S/\partial\omega\partial T,$ we have $(\partial
^{2}U/\partial\omega\partial T)/T=(\partial U/\partial T)/\omega+(\partial
^{2}U/\partial T\partial\omega)/T-[(U/\omega)+(\partial U/\partial
\omega)]/T^{2}$ or $0=(\partial U/\partial T)/\omega-[(U/\omega)+(\partial
U/\partial\omega)]/T^{2}.$ \ The general solution of this equation is%

\begin{equation}
U=f(T/\omega)\omega=\left\langle J\right\rangle \omega\label{U}%
\end{equation}
where $f\left(  T/\omega\right)  $ is an unknown function which corresponds to
the average value $\left\langle J\right\rangle $ of the action variable of the
mode. \ When applied to thermal radiation, the result obtained here purely
from thermodynamics corresponds to the familiar Wien displacement law of
classical physics.

\subsubsection{Possibility of Zero-Point Radiation}

The energy expression (\ref{U}) for an electromagnetic radiation mode (or for
a harmonic oscillator) in thermal equilibrium allows two limits which make the
energy independent from one of its two thermodynamics variables. \ When the
temperature $T$ becomes very large, $T>>\omega,$ so that the argument of the
function $f\left(  T/\omega\right)  $ is large, the average energy $U$\ of the
mode becomes independent of $\omega$ provided $f(T/\omega)\rightarrow
const_{1}\times T/\omega$ so that%

\begin{equation}
U=f(T/\omega)\omega\rightarrow const_{1}\times(T/\omega)\times\omega
=const_{1}\times T\text{ \ \ for \ }T/\omega>>1.\label{Uh}%
\end{equation}
This is the familiar high-temperature limit where we expect to recover the
Rayleigh-Jeans equipartition limit. \ Therefore we choose this constant as
$const_{1}=k_{B}$ corresponding to Boltzmann's constant. \ With this choice,
our thermal radiation now goes over to the Rayleigh-Jeans limit for high
temperature or low frequency.

In the other limit of small temperature, $T<<\omega,$ the dependence on
temperature is eliminated provided $f(T/\omega)\rightarrow const_{2},$ so
that
\begin{equation}
U=f(T/\omega)\omega\rightarrow const_{2}\times\omega\text{ \ \ for \ }%
T/\omega<<1.\label{Ul}%
\end{equation}
At this point, any theoretical description of thermal radiation must make a
choice. \ If we choose this second constant to vanish, $const_{2}=0,$ then
this limit does not force us to introduce any constant beyond Boltzmann's
constant which entered for the high-temperature limit of thermal radiation.
\ On the other hand, if we choose a non-zero value for this constant,
$const_{2}\neq0,$ then we are introducing a second constant into the theory of
thermal radiation, which constant has different dimensions from those of
Boltzmann's constant. \ The units of this new constant $const_{2}$ correspond
to \textit{energy} times \textit{length}. \ Furthermore, the choice of a
non-zero value for this constant means that at temperature $T=0,$ there is
random, temperature-independent radiation present in the system. \ This random
radiation which exists at temperature $T=0$ is classical electromagnetic
zero-point radiation. \ 

We emphasize that thermodynamics allows classical zero-point radiation within
classical physics. \ The physicists of the early 20th century were not
familiar with the idea of classical zero-point radiation, and so they made the
choice $const_{2}=0$ which excluded the possibility of classical zero-point
radiation. \ In his monograph on classical electron theory, Lorentz makes the
explicit assumption that there is no radiation present at $T=0.$\cite{Lorentz}
\ Today, we know that the exclusion of classical zero-point radiation is a
poor choice. \ However, the current textbooks of modern physics continue to
present only the outdated, century-old classical view.

Once the possibility of classical zero-point radiation is introduced into
classical theory, one looks for other phenomena where the zero-point radiation
will play a crucial role. \ In particular, the (Casimir) force between two
uncharged conducting parallel plates will be influenced by the presence of
classical electromagnetic zero-point radiation.\cite{Rev1} \ By comparing
theoretical calculations with experiments, one finds that the scale constant
for classical zero-point radiation appearing in Eq. (\ref{Ul}) must take the
value $const_{2}=1.05\times10^{-34}$Joule-sec. \ However, this value
corresponds to the value of a familiar constant in physics; it corresponds to
the value $\hbar/2$ where $\hbar$ is Planck's constant. \ Thus in order to
account for the experimentally observed Casimir forces between parallel
plates, the scale of classical zero-point radiation must be such that
$const_{2}=\hbar/2,$ and for each normal mode, the average energy becomes
\begin{equation}
U=f(T/\omega)\rightarrow(\hbar/2)\omega\text{ \ \ for }T\rightarrow
0.\label{zpr}%
\end{equation}

We emphasize that Planck's constant enters classical electromagnetic theory as
the scale factor in classical electromagnetic zero-point radiation. \ There is
no connection whatsoever to any idea of quanta. \ Many students are misled by
the textbooks of modern physics and regard Planck's constant as a "quantum
constant."\cite{essay} \ This is a completely misleading idea. \ A physical
constant is a numerical value associated with certain aspects of nature; the
constant may appear in several different theories, just as Cavendish's
constant G appears in both Newtonian physics and also in general relativity.
\ Here we emphasize that Planck's constant $\hbar$ appears in both classical
and quantum theories.

\section{Role of Relativity}

\subsubsection{Nonrelativistic Classical Physics Gives the Rayleigh-Jeans
Spectrum}

Today textbooks of modern physics present the Rayleigh-Jeans spectrum as
though it were the unique result of classical physics.\cite{texts}
\ Specifically, the equipartition-theorem result of \textit{nonrelativistic}
classical statistical mechanics is chosen as the average energy of each normal
mode of a harmonic oscillator and therefore for each normal mode of classical
thermal radiation. \ This energy-equipartition choice correponds to the
Rayleigh-Jeans result%
\begin{equation}
U_{RJ}=k_{B}T\label{RJ}%
\end{equation}
for each normal mode of radiation. \ As we have seen above, this choice is not
forced by thermodynamics. \ Rather it is the use \textit{nonrelativistic
}classical physics, either from \textit{nonrelativistic} statistical mechanics
or from \textit{nonrelativistic} scatters, which leads consistently to the
Rayleigh-Jeans result.

Indeed the importance of \textit{nonrelativistic} physics is evident in the
scattering calculations for random classical radiation. \ Clearly thermal
radiation should be stable under scattering by charged mechanical systems.
\ This stability is intrinsic to the idea of thermal equilibrium. \ A small
harmonic oscillator (dipole oscillator) scatters radiation without changing
the frequency of the incident radiation.\cite{Rev3} \ Thus a dipole oscillator
will enforce the isotropic nature of the radiation spectrum but will not
enforce any particular radiation spectrum for random classical radiation.
\ However, a \textit{nonlinear} dipole oscillator will indeed force an
equilibrium spectrum of random radiation; the equilibrium spectrum it enforces
is the Rayleigh-Jeans spectrum.\cite{nonlin} \ Indeed, the classical
zero-point radiation spectrum $U=(\hbar/2)\omega$ is unstable under scattering
by a \textit{nonrelativistic} nonlinear oscillator; the nonrelativistic
scatterer pushes the zero-point spectrum toward the Rayleigh-Jeans spectrum.

During the years around 1910, the question of the blackbody radiation spectrum
was discussed by the leading physicists of the day. \ It was suggested that
classical mechanics contained no constant comparable to Planck's constant
$\hbar$ appearing in the Planck blackbody radiation spectrum, and therefore
presumably classical physics was incapable of producing the Planck
spectrum.\cite{Lorentz2}

\subsubsection{Relativistic Classical Physics Gives the Planck Spectrum}

In Eq. (\ref{zpr}) above, we pointed out that the thermodynamics of classical
thermal radiation allows the possibility of classical electromagnetic
zero-point radiation. \ Indeed, if one accepts classical zero-point radiation,
then it turns out that the smoothest interpolation between the low-temperature
zero-point radiation limit and the high-temperature equipartition limit is
precisely the Planck spectrum including zero-point radiation\cite{SHO2} \ The
energy per normal mode for the Planck spectrum including zero-point radiation
is given by%
\begin{equation}
U_{P}=(\hbar\omega/2)\coth(\hbar\omega/(2k_{B}T)=\hbar\omega/2+\hbar
\omega(\exp[\hbar\omega/k_{B}T]-1)^{-1}\label{P}%
\end{equation}
\ Clearly one may wonder how this smooth-interpolation result fits with the
radiation scattering calculations.

It turns out that the spectrum of random classical electromagnetic zero-point
radiation in Eq. (\ref{zpr}) is invariant under adiabatic transformation.
\ The spectrum is also Lorentz invariant, scale invariant, and conformal
invariant.\cite{M}\cite{B-rel} \ Indeed the correlation functions of classical
zero-point radiation involve no preferred length, time, velocity, or
coordinate frame. \ The correlations involve only the geodesic separation
between the spacetime points where correlations are evaluated.\cite{geo} \ 

It is the Lorentz invariance of zero-point radiation which we wish to
emphasize here. \ Today physicists believe that all physics is intrinsically
relativistic; relativity is a meta theory to which all fundamental theories
should conform. Thus the use of nonrelativistic scatterers is at odds with the
Lorentz invariance of classical zero-point radiation and is also at odds with
our expectation that physics should be relativistic.

Suppose that we insist that the charged mechanical scatterers of thermal
radiation are \textit{relativistic} scatterers. \ In this case, we encounter
the relativistic conservation law\cite{conserv} and also the no-interaction
theorem of Currie, Jordan, and Sudershan.\cite{CJS} \ The only relativistic
interactions between particles involve either point collisions or else
interaction through a complete relativistic field theory. \ This theorem
reminds us why elementary treatments of relativity deal with particle point
collisions and never with particle interactions through a potential. The only
potential which has been extended to a complete relativistic field theory is
the Coulomb potential which is extended to charged particle interactions
within classical electrodynamics. \ 

Thus all the scattering calculations by \textit{nonrelativistic} dipole
systems are suspect. \ The one simple \textit{relativistic} scattering system
corresponds essentially to a hydrogen atom. \ And for the Coulomb potential
extended to relativistic electrodynamics, nature indeed includes a fundamental
mechanical constant with the units of \textit{energy} times \textit{time},
namely $e^{2}/c$, which is common to all interacting charged particles.  The
relativistic hydrogen atom has all the properties which suggest the
possibility of equilibrium with classical zero-point radiation and thermal
radiation at the Planck spectrum.\cite{scale} \ \ Indeed, in Goldstein's
mechanics text,\cite{Goldstein} we find the relativistic energy $\mathcal{E}$
for a particle of mass $m$ in a Coulomb potential $Ze^{2}/r$ in terms of
action-angle variables $J_{2}$ and $J_{3}$ takes the form\cite{2pi}
\begin{equation}
\frac{\mathcal{E}}{mc^{2}}=\left(  1+\left[  \frac{J_{3}c}{Ze^{2}}-\frac
{J_{2}c}{Ze^{2}}+\left\{  \left(  \frac{J_{2}c}{Ze^{2}}\right)  ^{2}%
-1\right\}  ^{1/2}\right]  ^{-2}\right)  ^{-1/2}\label{Coul}%
\end{equation}
\ We notice immediately that the constant $Ze^{2}/c$ is a crucial parameter
for the system. \ If $J_{2}<Ze^{2}/c,$ then the energy expression in Eq.
(\ref{Coul}) is no longer valid because it involves the square-root of \ a
negative quantity. \ This is a reminder that the constant $e^{2}/c$ is a
crucial parameter in \textit{relativistic} classical physics and that the
orbits of relativistic motion are quite different from the nonrelativistic
elliptical orbits. \ Indeed, if $J_{2}<Ze^{2}/c,$ then (neglecting radiation
emission) the relativistic mechanical trajectories plunge into the Coulomb
center while conserving energy and angular momentum.\cite{relM} \ Relativistic
systems can be quite different from nonrelativistic systems. \ 

\subsubsection{Relativistic Physics and Gravitation}

The enormous difference between relativistic and nonrelativistic thermodynamic
systems is immediately evident if we consider a thermodynamic system in
gravity. \ Indeed, Boltzmann derived the Maxwell velocity distribution for
nonrelativistic particles in thermal equilibrium by considering thermal
equilibrium for particles under a gravitational field. \ In nonrelativistic
physics, gravity couples only to the masses of particles; it does not couple
to kinetic energy or potential energy. \ Accordingly, the pressure of a
nonrelativistic system in thermal equilibrium under gravity reflects the
changing particle density with height, while the temperature remains constant
throughout the system. \ By considering pressure equilibrium and
nonrelativistic particle motion in the vertical direction, Boltzmann was led
to the Maxwell velocity distribution.\cite{Boltz} \ 

The situation in relativistic physics is quite different. \ In relativistic
physics, gravity couples to all energy, including both kinetic energy and
energy in the electromagnetic fields. \ Thus if we consider flat spacetime but
go to a Rindler frame undergoing time-independent proper acceleration, then
neither the equivalence-principle gravitational field nor the temperature of a
thermal system in equilibrium can be constant with changing distance from the
event horizon. \ At spatial coordinates which are lower in the gravitational
field, the temperature is higher so that $(g_{00})^{1/2}T=const,$ where
$g_{\mu\nu}$ corresponds to the metric tensor for the spacetime.\cite{Tol}
\ If we consider classical electromagnetic zero-point radiation to be present,
then the correlation in time at a given spatial coordinate reflects the
gravitational field at the point. \ By carrying out a scaling transformation
in time within a Rindler frame, one changes from thermal equilibrium at zero
temperature over to thermal equilibrium at a finite non-zero temperature.
\ Finally, by maintaining constant the local temperature as one moves ever
further from the event horizon, one can move to a region where the spacetime
metric becomes the Minkowski metric of an inertial frame while maintaining the
finite temperature spectrum. \ The thermal spectrum which one finds is exactly
the Planck spectrum.\cite{T-rel} \ Thus a relativistic treatment indeed gives
the Planck spectrum for thermal radiation. \ One notes that the derivation
requires relativistic behavior at every step of the analysis. \ 

By analyzing a nonrelativistic thermal situation under gravity, Boltzmann
derived the thermal distribution for particles. \ By analyzing the
relativistic situation for radiation, including zero-point radiation, under
gravity, one derives the Planck spectrum for thermal radiation. \ The
difference between the Rayleigh-Jeans spectrum and the Planck spectrum does
not represent a difference between classical and quantum physics. \ Rather the
difference between the Rayleigh-Jeans spectrum and the Planck spectrum for
thermal radiation within classical physics corresponds to the use of
nonrelativistic versus relativistic physics.

\section{Discussion}

In this short sketch, we have outlined the analysis of the blackbody radiation
spectrum from a modern classical point of view which includes both classical
zero-point radiation and notes the importance of relativity. \ The inclusion
of classical electromagnetic zero-point radiation will affect many aspects of
the physics of small objects where the strong zero-point radiation
fluctuations at high frequencies become important. \ Thus classical
electromagnetic zero-point radiation has been used to give detailed
calculations of Casimir forces, van der Waals forces, classical oscillator
behavior, specific heats of solids, and diamagnetism.\cite{Rev1} \ These
calculations give results in complete agreement with the calculations of
quantum physics. \ The presence of classical zero-point radiation also
transforms the old problem of atomic collapse where a planetary electron is
claimed to spiral into the nucleus as the electron accelerates and radiates
away its energy. \ In the presence of classical zero-point radiation, the
spiraling electron will perform an orbital Brownian motion where it absorbs
radiation from the random zero-point radiation field as well as emitting
radiation during acceleration. \ Numerical simulations show that the electron
in a Coulomb orbit does not collapse into the nucleus when classical
zero-point radiation is present.\cite{CZ}\cite{NL}

When one reads the most recent textbooks of modern physics with knowledge of
the current state of classical electromagnetic theory, one can only be struck
by the irony of these texts. \ Today modern physics texts invariably start
with a discussion of relativity. \ However, the texts then drop the subject
completely. \ The texts give no suggestion that the ideas of relativity may
have significance for the discussions of blackbody radiation and atomic
physics which follow. \ The texts contain many references to nonrelativistic
mechanics but not one mention of the no-interaction theorem of relativistic mechanics.

There are many close and fascinating connections between classical and quantum
physics. \ Relativistic classical electron theory is currently the best
classical approximation to quantum physics. \ Discussions which present an
outdated view of classical physics do a disservice to students who will become
the scientists of the future.


\begin{thebibliography}{99}
\bibitem {Lorentz2}In ref. 9, Lorentz returns repeatedly to the idea that
Planck's blackbody spectrum requires that some aspect of nature must be the
same for all mechanical systems. \ See, for example, Section 58 on p. 78. \ In
Section 75 on page 96, Lorentz notes the constancy of the
Wien-displacement-theorem quantity $T\lambda_{\max}.$ \ He remarks,
\textquotedblleft\lbrack If we can account for this constant], we may hope to
find in what manner the value of this constant is determined by some numerical
quantity that is the same for all ponderable bodies.\textquotedblright
\bibitem {2pi}Here we have chosen action-angle variables differing by a factor
of $1/(2\pi)$ from those of problem 28 of Goldstein's text and have written
the strength of the potential as $k=Ze^{2}.$
\end{thebibliography}
\end{document}